\documentclass[aps,pre,12pt,a4paper,nofootinbib]{revtex4-2}
\usepackage{epsfig}
\usepackage{graphicx}
\usepackage{amsmath,amssymb,color}
\usepackage[english]{babel}
\usepackage{natbib}
\usepackage{appendix}
\usepackage{caption}
\usepackage{subcaption}

\parskip=\medskipamount

\newcommand{\be}{\begin{equation}}
\newcommand{\ee}{\end{equation}}
\newcommand{\eq}[1]{(\ref{#1})}
\newcommand{\fig}[1]{Fig.~\ref{#1}}
\newcommand\disp{\displaystyle}

\newcommand{\eps}{\varepsilon}

\begin{document}

\title{KPZ-like scaling on a high-dimensional hypersphere}

\author{Daniil Fedotov$^1$ and Sergei Nechaev$^2$}

\affiliation{$^1$UFR Sciences, Universit\'e Paris Saclay, 91405 Orsay Cedex, France \\ 
$^2$LPTMS, CNRS -- Universit\'e Paris Saclay, 91405 Orsay Cedex, France}


\begin{abstract}

We consider the orientational diffusion controlled by the hyperspherical Laplacian, $\nabla^2_D$, on the surface of the $D$--dimensional hypersphere in the limit $D \to \infty$. We find that for stretched paths with lengths relatively short compared to the hypersphere's radius, the finite-size corrections in orientational correlations are controlled by the Kardar-Parisi-Zhang (KPZ) scaling exponent, $\gamma = 1/3$. In addition, we speculate about the topology of the orientational target space representing the surface of the hypersphere.

\end{abstract}

\maketitle

\section{Introduction}
\label{sect:01}

Scaling laws of fluctuations serve as a useful tool for examining the statistical behavior of systems. Two well-known laws, the ``Gaussian'' and the ``Tracy-Widom'', appear in a wide range of statistical systems. The familiar Gaussian law is generally associated with the central limit theorem, which applies to large sets of independent random variables. In contrast, the Tracy-Widom (TW) law \cite{tracy} which is manifested in the Kardar-Parisi-Zhang (KPZ) scaling \cite{kpz,kpz-review} often arises in the extreme statistics of large sets of correlated random variables.

Many (1+1)-dimensional and two-dimensional models exhibit the critical behavior associated with the 1$D$ Kardar-Parisi-Zhang (KPZ) universality class. Interest in these systems was inspired by the (1+1)$D$ model proposed by H. Spohn and P. Ferrari in \cite{ferrari}, where they explored the statistics of 1D directed random walks that avoid a semicircular constraint. As stated by the authors in \cite{ferrari}, their motivation was as follows. A set of $n$ one-dimensional directed ``vicious walks'' connected at their endpoints, represent the ensemble of world lines of free fermions in 1D and the extreme (top) world line shares the Tracy-Widom distribution \cite{tracy,tracy-review}. In \cite{spohn-pr1, spohn-pr2}, the authors defined the average position of the top line and studied its fluctuations. In this framework, the walks below the top line act as a ``mean field'' representing the bulk, which pushes the top line towards an equilibrium position. The fluctuations around this position differ from those of a free random walk without the bulk's influence. By modeling the bulk's effect with a semicircle, the Spohn-Ferrari model is obtained, in which the 1$D$ directed random walk remains above the semicircle, with the region inside it inaccessible to the path. In \cite{ferrari}, the authors confirmed that this system exhibits a KPZ critical exponent.

The effective interaction between a path and an extended void was studied in detail in \cite{nech-val-gor1, polovnikov, baruch1, baruch2, baruch3, shlosman, nech-val-gor2} within the context of the 2$D$ extension of the (1+1)D Ferrari-Spohn problem. When reformulated in polymer terms, the problem can be described as follows: a portion of a polymer's trajectory near the boundary of a convex void undergoes a transition between different fluctuation regimes, depending on the curvature of the boundary. This behavior was interpreted in \cite{baruch1, baruch2} as ``shadowing'' of a path by the convex, impermeable boundary of a disc, against which the path leans. In this framework, the key aspect is the consideration of a microcanonical ensemble of fluctuating paths of fixed length, $N$. The condition $N = cR$, where $c$ is a constant, is imposed on paths whose ends are fixed at the opposite extremities of a diameter of the convex void (a circle of radius $R$). By varying the control parameter $c$, one observes a transition from Kardar-Parisi-Zhang (KPZ) scaling to Gaussian behavior. This transition is driven by fixing the path length, which forces the trajectories to conform to the disc boundary. The parameter $c$ controls the strength of this ``pinching force'' and effectively represents the curvature of the disc's boundary relative to the path's length.

An alternative perspective on exactly solvable one-body statistical models exhibiting KPZ fluctuation behavior has been proposed in \cite{nech-val-gor1,nech-val-gor2,nech-val-gor3}. These works discuss the problem of path counting on a nonhomogeneous Cayley tree, where the branching depends linearly on the tree's generation. In the thermodynamic limit, KPZ-like scaling for random walks on such nonhomogeneous trees (NT) is observed. The model for path counting on NT can be considered as a mean-field approximation of the Dumitriu-Edelman framework \cite{edelman} of random matrix theory since the Hermite polynomials are the eigenfunctions of the transfer matrix for the problem of paths counting on NT.

The question addressed in the present work has some mathematical intersections with the problem of path counting on nonhomogeneous trees, though it stems from an entirely different physical context. Specifically, our focus is on the statistics of random walks in homogeneous high-dimensional space, or more precisely, in infinite-dimensional space. We aim to demonstrate that, in hyperspherical coordinates, under a specific scaling of the random walk parameters, the probability distribution of directional correlations along the path has a KPZ-type scaling.

The paper is structured as follows. In Section \ref{sect:02}, we review the basic geometrical properties of high-dimensional spaces and discuss the solution of the diffusion equation in hyperspherical $N$-dimensional coordinates paying attention to the limit $N \to \infty$. In Section \ref{sect:03}, we analyze the resulting analytic expression, derive the KPZ-like stretched exponential behavior of directional correlations along the path, and present a numerical implementation of the model under consideration. In Section \ref{sect:04} we summarize obtained results, speculate about the topology of the underlying target space, and discuss possible connections with other models possessing KPZ-like scaling.

\section{Angular diffusion in $D$-dimensional hyperspherical coordinates}
\label{sect:02}

There is a common belief that the high-dimensional space is topologically similar to a tree. This belief is grounded in the straightforward extrapolation of the local structure of the high-dimensional space (i.e. absence of short cycles) onto its general topology. However the high-dimensional space holds paradoxes and one of the most known ones emerges when computing the volume, $V_D(R)$, and the surface, $S_D(R)$, of the $D$-dimensional sphere of radius $R$:
\be
\left\{\begin{array}{lll}
\disp V_{2D}(R) = \frac{\pi^D}{D!}R^{2D}, \quad &  \disp S_{2D-1}(R) = \frac{2 \pi^D}{(D-1)!}R^{2D-1} \quad & \mbox{for $D=2,4,6,...$} \medskip \\
\disp V_{2D+1}(R) =  \frac{2^{2D+1}D!}{(2D+1)!}R^{2D+1}, \quad & \disp S_{2D}(R) = \frac{2^{2D+1} D!\pi^D}{(2D)!}R^{2D} \quad & \mbox{for $D=1,3,5,...$}
\end{array}
\right.
\label{eq:01}
\ee
Considering the quotient $V_D(R+\eps)/V_D(R)$, we get for fixed $R$ and $0<\eps\ll R$ in the limit $D\to\infty$:
\be
\left.\frac{V_{D}(R+\eps)}{V_D(R)}\right|_{D\gg 1} \to \left.\left(1+\frac{\eps}{R}\right)^D\right|_{D\to \infty} \to \infty
\label{eq:02}
\ee
Since $V_D(R+\eps) = V_D(R) + Vh_D$, where $Vh_D$ is the volume of a hyperspherical layer, we immediately conclude that $\lim_{D\to\infty} Vh_D/V_D \to\infty$ and the whole volume of the $D$-dimensional sphere is concentrated in its hyperspherical surface layer. Thus, the hyperspherical coordinate system is most suitable for analyzing the structure of high-dimensional spaces.

It is legitimate to ask what topological structure the $D$-dimensional metric space in hyperspherical coordinates has in the limit $D \to \infty$. One way to explore the structure of this metric space is by examining its spectral properties through the solution of the corresponding eigenvalue problem. This question has been repeatedly addressed in the literature (see, for example, \cite{Vilenkin, Radoslaw, Jing}). Here, we reproduce some of the results necessary for subsequent analysis.

Consider the diffusion equation for the probability density $W(r,\pmb{\varphi},t)$ in the full $D$-dimensional space in hyperspherical coordinates $(r,\pmb{\varphi}) = \left(r, \{\varphi_1, \varphi_2,...,\varphi_{D-1}\}\right)$. For $D\gg 1$ the contribution from volumic part is negligible compared to the one on the surface of the $D\gg 1$--dimensional hypersphere (see \eq{eq:01}) and in the first approximation at $D\to\infty$ we may disregard the diffusion in the bulk of the hypersphere leaving only the diffusion on the hyperspherical surface ``along'' the angular coordinates: 
\be
\begin{cases}
\disp \partial_t W(\pmb{\varphi},t) = a^2 \Delta W(\pmb{\varphi},t) \medskip \\
\disp W(\pmb{\varphi}, t)\big|_{t=0} =\frac{\delta^{S_{D-1}}\left(\pmb{\varphi} \,\pmb{\varphi}'\right)}{R^{D-1} A_{D-1}} 
\end{cases}
\label{eq:03}
\ee
where by $a^2$ we have denoted the diffusion coefficient, $R$ is the radius of the $D$-dimensional hypersphere, $\delta^{S_{D-1}}(...)$ is the $\delta$-function on the surface $S_{D-1}$ of $D$-dimensional hypersphere, and $\pmb{\varphi}$ is the angular coordinate of some point of path on $S_{D-1}$. The notation $\pmb{\varphi}\, \pmb{\varphi}'$ denotes $\cos\widehat{\pmb{\varphi}\, \pmb{\varphi}'}$ where $\widehat{\pmb{\varphi}\, \pmb{\varphi}'}$ is the angle between two rays emitted from the center of the hypersphere to the points $\pmb{\varphi}$ and $\pmb{\varphi}'$ on $S_{D-1}$. In what follows we set for simplicity $a^2=1$ and consider the diffusion in the dimensionless units on the surface of $D$-dimensional sphere of radius $R$. 

The Green function  $W(\pmb{\varphi}, t | \pmb{\varphi}', 0)$ can be thought as the probability density of the ensemble of random paths of length $t$ having the angular distance $\cos\widehat{\pmb{\varphi}\, \pmb{\varphi}'}$ on the hypersphere $S_D$. The solution to \eq{eq:03} is well known -- see, for example, \cite{Caillol, Vilenkin, Radoslaw, Jing} -- and it can be written as a series over hypersherical eigenfunctions, which are the Gegenbauer polynomials, $C_k^{D/2-1}$, weighted with the corresponding eigenvalues $\lambda_k$ :
\be
W(\pmb{\varphi}, t | \pmb{\varphi}', 0)=\frac{1}{R^{D-1} A_{D-1}}\sum_{k=0}^{\infty} \frac{2k+D-2}{D-2} C_k^{D/2-1}\left(\pmb{\varphi}\, \pmb{\varphi}'\right) e^{-\lambda_k t/R^2}, \quad k=0,1,2,\ldots
\label{eq:04}
\ee
where $\pmb{\varphi}$ and $\pmb{\varphi}'$ are two vectors corresponding to extremities of the path of length $t$ on hypersphere $S_D$, and 
\be
\lambda_k=k(k+D-2) 
\label{eq:05}
\ee

The Gegenbauer polynomials, $C_k^{\alpha}(x)$ can be defined for example, as the coefficients of the generating function expansion, cite{Gegenbauer}:
\be
(1 - 2x u + u^2)^{-\alpha} = \sum_{n=0}^{\infty} C_k^{\alpha}(x) u^k  
\label{eq:06}
\ee
where $\alpha=D/2-1$. The polynomials $C_k^{\alpha}(x)$ as a function of $x$ are defined on the interval $x\in [-1,1]$ and have obvious symmetric property: $C_k^{\alpha}(-x) =(-1)^k C_k^{\alpha}(x)$. 

\section{KPZ-like scaling of angular diffusion on a surface of a $D\gg 1$-dimensional  hypersphere}
\label{sect:03}

It has been established in \cite{infty1, infty2, infty3} that in an infinitely-dimensional space the appropriately normalized eigenvectors of the the angular part of a spherically-symmetric Laplacian $\nabla^2_{\infty}$, i.e. the Gegenbauer hyperspherical functions, converge to the Hermite polynomials. Namely, in the limit $D\to \infty$ (and hence, for $\alpha\to\infty$) the following relation holds \cite{infty3,temme}:
\be
\lim_{\alpha\to\infty}\alpha^{-k/2} C_k^{\alpha}(y/\sqrt{\alpha}) = \frac{1}{k!}H_k(y)
\label{eq:07}
\ee
where $H_k(y)$ is the Hermite polynomial
\be
H_k(y)=k!\sum_{m=0}^{\infty}\frac{(-1)^m}{m!(k-2m)!}(2y)^{k-2m}
\label{eq:08}
\ee
and $y=\sqrt{\tfrac{D}{2}-1}\;\pmb{\varphi}\, \pmb{\varphi}'\equiv \sqrt{\tfrac{D}{2}-1}\; \cos\widehat{\pmb{\varphi}\, \pmb{\varphi}'}$. Thus, one has from \eq{eq:07}--\eq{eq:08}:
\begin{multline}
\left.\left(\tfrac{D}{2}-1\right)^{-k/2} C_k^{\frac{D}{2}-1}\left(y/\sqrt{\tfrac{D}{2}-1}\right)\right|_{\frac{D}{2}-1 \gg 1}\to \frac{1}{k!} H_k(y) = \medskip \\ \frac{1}{k!} H_k \left(\sqrt{\tfrac{D}{2}-1}\;\pmb{\varphi}\, \pmb{\varphi}'\right) = \frac{1}{k!} H_k \left(\sqrt{D-2}\;\pmb{\varphi}\, \pmb{\varphi}'/\sqrt{2}\right) = \frac{2^{k/2}}{k!}\mathcal{H}_k\left(\sqrt{D-2}\, \pmb{\varphi}\, \pmb{\varphi}'\right) 
\label{eq:09}
\end{multline}
where  ${\cal H}_k(y)$ are the probabilistic Hermite polynomials:
\be
{\cal H}_k(y)=(-1)^k e^{\frac{y^2}{2}}{\frac{d^k} {dy^k}}e^{-{\frac{y^2}{2}}}; \qquad {\cal H}_k(y)=2^{-k/2}H_k(y/\sqrt{2)}
\label{eq:10}
\ee
The polynomials ${\cal H}_k(y)$ are defined on the whole axis $y\in ]-\infty, \infty[$ and have the same symmetry as Gegenbauer polynomials, i.e. ${\cal H}_k(-y) =(-1)^k {\cal H}_k(y)$. In what follows we shall be interested in the behavior of ${\cal H}_k(y)$ on the positive ray $y\ge 0$ only. 

The dominant contribution to the probability distribution $W(\pmb{\varphi},t |\pmb{\varphi}',0)$, as defined by the series in \eq{eq:04}, for long paths $t \gg 1$, arises from all values of $k$ up to $k=k_{max}$ such that $\lambda_{k_{max}} t/R^2\approx 1$. This sets the value of $k_{max}$ at which the series \eq{eq:04} can be cut. 

\subsection{Analysis of the different regimes of the series \eq{eq:04}}
\label{sect:03a}

Let begin with the consideration of very long paths, $t\gg R^2$. In that case $\lambda_k\to 0$ and the contribution to the series \eq{eq:04} (as it follows from \eq{eq:05}) comes from $k=0$ with the first nontrivial correction giving by $k=1$:
\be
W\left(\pmb{\varphi},t|\pmb{\varphi}',0\right) = \frac{1}{R^{D-1}A_{D-1}}\left(1+\frac{2D}{D-2} \cos \widehat{\pmb{\varphi}\, \pmb{\varphi}'}\; e^{-(D-1) t/R^2}\right)
\label{eq:11}
\ee
As one sees, the probability $W\left(\pmb{\varphi},t|\pmb{\varphi}',0\right)$ demonstrates the standard exponential decay in time at $t\gg R^2/(D-2)$.

A much richer behavior occurs if one considers relatively short trajectories compared to $R^2$, i.e. when we are in the situation $t \ll  R^2$. In that case the series expansion \eq{eq:04} saturates at sufficiently large $k=k_{\rm max}$: i.e. the combination $\lambda_{k_{\rm max}} t / R^2$ becomes of order of unity for large $k_{\rm max}$ and all terms in the series \eq{eq:04} up to $k_{\rm max}$ matter. Reverting \eq{eq:05}, we get the following dependence $k(\lambda_k)$
\be
k=k_{\rm max}=\frac{1}{2}\left(\sqrt{4\lambda_k+(D-2)^2}-(D-2)\right)
\label{eq:12}
\ee
At $\lambda_k\gg 1$ one also has $k_{\rm max}\gg 1$. In that case, all terms in the series \eq{eq:04} from $k=0$ till $k = \lfloor k_{\rm max} \rfloor$, where $k_{\rm max}$ is given by \eq{eq:12}, contribute to $W\left(\pmb{\varphi},t|\pmb{\varphi}',0\right)$.

Suppose we are now interested in the behavior of the probability distribution $W\left(\pmb{\varphi},t |\pmb{\varphi}',0\right)$ when $\cos \widehat{\pmb{\varphi}\,\pmb{\varphi}'} \approx 1$, meaning that all paths are ``stretched'' along the angular coordinates on the hypersurface $S_{D-1}$. In this regime, we expect the distribution to exhibit a localized peak around $\pmb{\varphi} \approx \pmb{\varphi}'$, with a small spread around. Estimation of the width of this spread is our main concern here.

\begin{figure}[ht]
\centering
\includegraphics[width=0.95\linewidth]{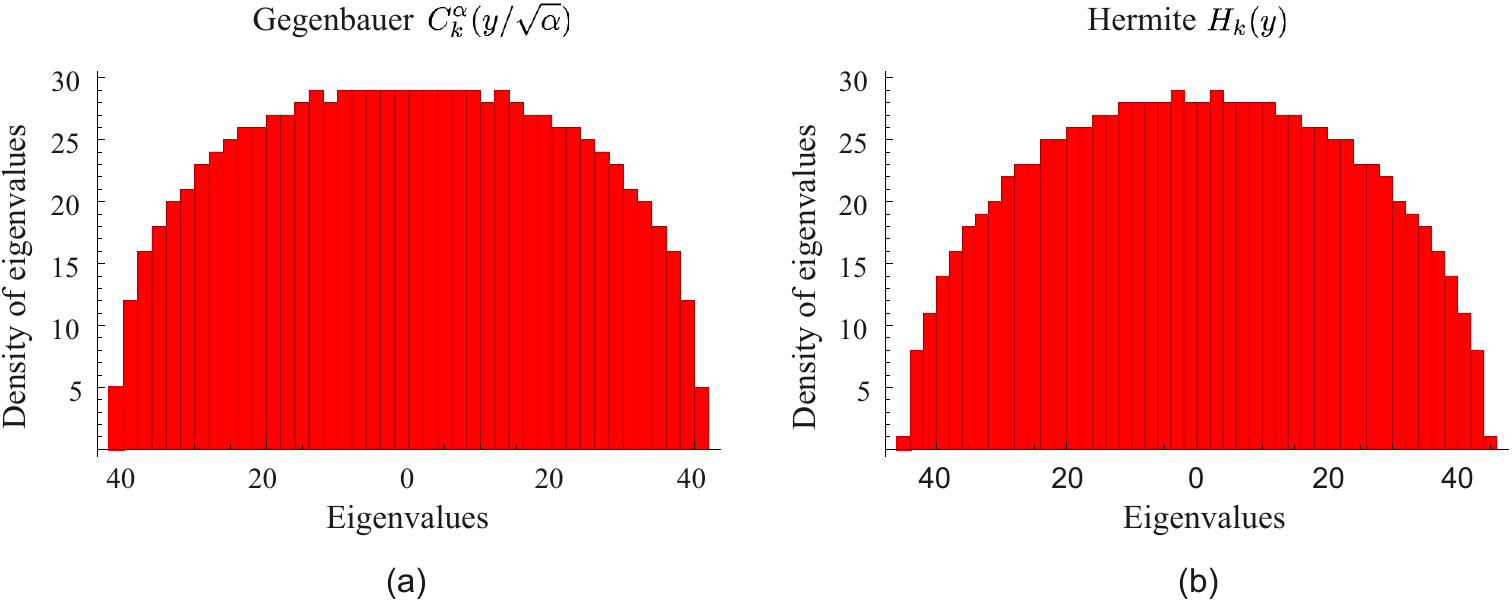}
\caption{Distribution of eigenvalues: (a) of Gegenbauer polynomial $C_k^{\alpha}\left(y/\sqrt{\alpha}\right)$ where $\alpha = \frac{D}{2}-1$, for $k=1000$ and $D=20\,000$; (b) of Hermite polynomials, $H_k(y)$ for $k=1000$. The spectral boundary is approximately $|\lambda_{\rm max}|\approx \sqrt{2k}=\sqrt{2\times 1000}\approx 44.7$.}
\label{fig:01}
\end{figure}

Recall that the distribution of eigenvalues of probabilistic Hermite polynomials, ${\cal H}_k(y)$, defined by \eq{eq:10}, shares the Wigner semicircle law for random matrix ensembles \cite{hermite-semi}, and all eigenvalues of ${\cal H}_k(y)$ lie within the interval $y \in \left[-2\sqrt{k},\, 2\sqrt{k}\right]$. Since the argument of $H_k(y)$ is rescaled by a factor $\sqrt{2}$ with respect to the one of ${\cal H}_k(y)$, the distribution of eigenvalues of $H_k(y)$ is given by the rescaled Wigner semicircle and lie on the interval $y \in \left[-\sqrt{2k},\, \sqrt{2k}\right]$. In \fig{fig:01}a we have plotted the distribution of eigenvalues of the Gegenbauer polynomials, $C_k^{\frac{D}{2}-1}\left(y/\sqrt{\tfrac{D}{2}-1}\right)$, for $k=1000$ and $D=20\,000$, while the distribution of eigenvalues the Hermite polynomials, $H_k(y)$ for $k=1000$ is shown in \fig{fig:01}b. These distributions are nearly identical, sharing the rescaled Wigner semicircle with all eigenvalues belonging to the interval $\left[-\sqrt{2k},\, \sqrt{2k}\right]$. Since the distribution of eigenvalues is symmetric, we consider only the positive part of the spectrum, $y \ge 0$. 

In the vicinity of the spectral edge $y_{\rm max}=2\sqrt{k}$ the asymptotic behavior of Hermite polynomials has been analyzed in \cite{Dominici} where an expansion in terms of the Airy function has been found:
\begin{multline}
H_k\left(\sqrt{\frac{D-2}{2}}\;\pmb{\varphi}\, \pmb{\varphi}'\right) \approx \sqrt{2 \pi} \exp \left(\frac{k \ln (2 k)}{2}-\frac{3 k}{2}+\sqrt{D-2}\;\pmb{\varphi}\, \pmb{\varphi}\,\sqrt{k}\right) k^{1 / 6} \times \\ \mathrm{Ai}\left(\frac{\sqrt{D-2}\;\pmb{\varphi}\, \pmb{\varphi}'-2 \sqrt{k}}{k^{-1 / 6}}\right)
\label{eq:13}
\end{multline}
Recall that the Airy function is defined as follows: 
$$
\disp {\rm Ai}(z)=\frac{1}{\pi} \int_{0}^{\infty} \cos(\xi^3/3+\xi z)\, d\xi
$$ 
The function ${\rm Ai}(z)$ has zeros at points $...<a_i<...<a_2<a_1$ ($a_i<0$ for all $i$). In the vicinity of the value $2\sqrt{k}$ (for $k\gg 1$), the argument of the Airy function, $\sqrt{D-2}\;\pmb{\varphi}\, \pmb{\varphi}'$, has the following ``finite-size'' expansion
\be
\sqrt{D-2}\;\pmb{\varphi}\, \pmb{\varphi}' \approx 2\sqrt{k} + a_1 k^{-1/6}
\label{eq:14}
\ee
where $a_1\approx -2.3381$ is the first zero of Airy function.

To see the manifestation of the Airy-type behavior \eq{eq:13}--\eq{eq:14}, two conditions should be fulfilled: 
\begin{itemize}
\item[(i)] The combination $\sqrt{D-2}\;\pmb{\varphi}\, \pmb{\varphi}'$ should stay near the value $2\sqrt{k_{\rm max}}$, which implies the condition 
\be
\pmb{\varphi}\,\pmb{\varphi}' \equiv \cos \widehat{\pmb{\varphi}\,\pmb{\varphi}'} = \frac{2\sqrt{k_{\rm max}}}{\sqrt{D-2}} \approx 1 \quad \Rightarrow \quad k_{\rm max} = \frac{1}{4} (D-2)
\label{eq:15}
\ee
\item[(ii)] The main contribution to the series \eq{eq:04} is given by the last term only, with $k=k_{\rm max}$, at which this series is cut. 
\end{itemize}
Here, we focus on condition (i), while the numerical verification of condition (ii) is addressed in Subsection \ref{sect:03b}.

Eq.\eq{eq:15} sets the value $k_{\rm max}$ in the series \eq{eq:05} and this value should be identical to $k_{\rm max}$ defined in \eq{eq:12}. Comparing these two expressions, we get the equation
\be
\frac{1}{4}(D-2) = \frac{1}{2}\left(\sqrt{4\lambda_k+(D-2)^2}-(D-2)\right)
\label{eq:16}
\ee
from which it follows that $\lambda_{\rm max} = \tfrac{5}{16}(D-2)^2$. Remembering that $\lambda_{\rm max} \approx R^2/t$, we arrive at the estimate on the paths lengths, $t$, at which we could expect the nontrivial scaling behavior of the distribution function $W\left(\pmb{\varphi},t |\pmb{\varphi}',0\right)$:
\be
t \lesssim \frac{16}{5} \left(\frac{R}{D-2}\right)^2; \qquad (R\gg D\gg 1)
\label{eq:17}
\ee

Collecting \eq{eq:15} and \eq{eq:17} and plugging them into \eq{eq:14} we arrive at the following finite-size correction to the directional correlations of stretched random walk of length $t$ on the surface of the high-dimensional hypersphere:
\be
\pmb{\varphi}\,\pmb{\varphi}' \equiv \cos \widehat{\pmb{\varphi}\,\pmb{\varphi}'} \approx 1 + b \left(\frac{t}{R^2}\right)^{1/3}, \quad (b\le 0)
\label{eq:18}
\ee
where $b=\left(\tfrac{5}{8}\right)^{1/3}a_1 \approx -1.9991$.

\subsection{Numerical verification of condition (ii)}
\label{sect:03b}

Derivation described in previous Subsection \ref{sect:03a} is based on the assumption that the in the series \eq{eq:04} the term $k=k_{\rm max}=\frac{1}{2}\left(\sqrt{4\lambda_k+(D-2)^2}-(D-2)\right)$ provides the main contribution to the sum. Here we verify numerically that it is always possible to chose such $R$ which ensures this condition to be true. To do that, we fix $t$ and, for each $D$, determine the required $R$.

The Gegenbauer polynomial, $C_k^{(\alpha)}(x)$, in the vicinity boundary values $x\to\pm 1$ has the following uniform asymptotic expansions \cite{uniform}:
\be
C_k^{(\alpha)}(x) \sim \frac{\Gamma(2\alpha + k)}{\Gamma(2\alpha)k!} (1 - x^2)^{-\alpha/2},
\label{eq:19}
\ee
For large $k$, in the vicinity of $x = \pm 1$, the Gegenbauer polynomials exhibit rapid growth. This behavior is attributed to the factor $(1-x^2)^{-\alpha/2}$, which tends to infinity as $x$ approaches $\pm 1$, especially for positive values of $\alpha$ (in our case $\alpha = D/2 - 1 \gg 1$).

\begin{figure}[ht]
\centering
\includegraphics[width=0.7\linewidth]{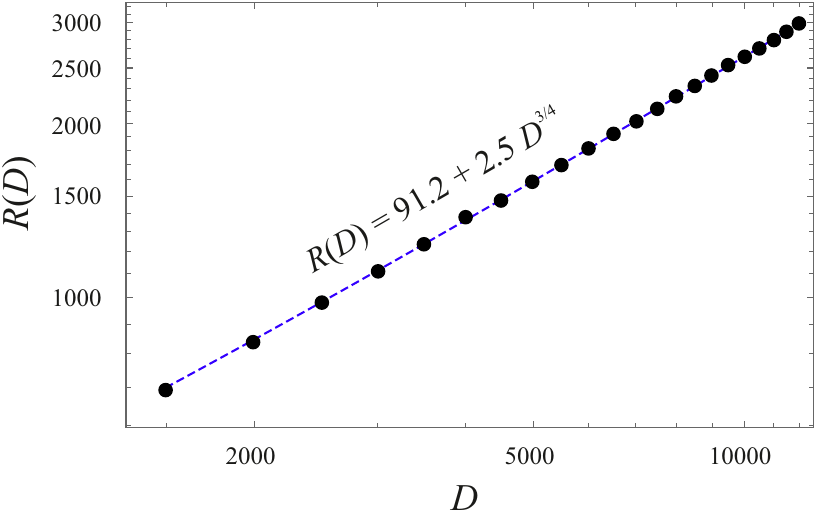}
\caption{The log-log plot illustrates the relationship $R(D)$ obtained in the numerical simulations (black dots) which sets the value of $R$ for each $D$ that ensures the validity of the condition (ii). The approximating curve (dashed blue) demonstrates the scaling dependence $R(D) \approx 91.2 + 2.5 D^{3/4}$.}
\label{fig:02}
\end{figure}

This implies that, in our case, it suffices to evaluate the values of $C_k^{(\alpha)}(x)$ at the endpoints of the interval $[-1,1]$. Using the symmetry property $C_k^{(\alpha)}(-x) = (-1)^k C_k^{(\alpha)}(x)$, we only need to consider the point $x=1$. Let us fix the dimension $D$. If the absolute value of the $k_{\text{max}}$th term at $x=1$ in the series \eq{eq:04} exceeds the sum of all other terms by an order of magnitude for a certain $R = R(D)$, we select this $R$ \footnote{If multiple such values exist, we choose the one for which the value of the Gegenbauer polynomial of the $k_{\text{max}}$th term at $x=1$ is maximal.}. Our numerical computations demonstrate that, for any $D$, it is always possible to find such an $R(D)$. Furthermore, we found that this $R(D)$ follows the relationship $R(D) \sim D^{3/4}$, as illustrated in \fig{fig:02}.

The \fig{fig:02} is plotted as follows: we consider the dimensionality $D$ within the range $D \in [1500, 12000]$ with a step of $\Delta D = 500$. For each value of $D$, we determine $R$ that satisfies condition (ii). For instance, when $D = 3000$, we use $R = 1112.96$. Notably, for this value of $R$, the $k_{\text{max}}$th term in the series consistently dominates the sum of the preceding $k_{\text{max}}-1$ terms, as required.

\section{Conclusion}
\label{sect:04}

We have demonstrated that accounting for orientational diffusion on the surface of a $D \gg 1$-dimensional hypersphere of radius $R$ leads to non-Gaussian KPZ-like finite-size corrections for relatively short trajectories of length $t$ compared to $R^2$. Specifically, these corrections, derived in \eq{eq:18}, are valid for $t \lesssim \frac{16}{5} \left(\frac{R}{D-2}\right)^2$ (see \eq{eq:17}). Our approach relies on two key components: 

1. The eigenfunctions of the Laplacian on the $D \gg 1$-dimensional hypersphere, the Gegenbauer polynomials, reduce to Hermite polynomials under proper normalization (see \eq{eq:07}). 

2. We analyze the ``stretched'' limit of orientational diffusion, where $\pmb{\varphi} \cdot \pmb{\varphi}' \equiv \cos \widehat{\pmb{\varphi}, \pmb{\varphi}'} \approx 1$ (see \eq{eq:13}--\eq{eq:15}). 

\begin{figure}[ht]
\centering
\includegraphics[width=0.6\textwidth]{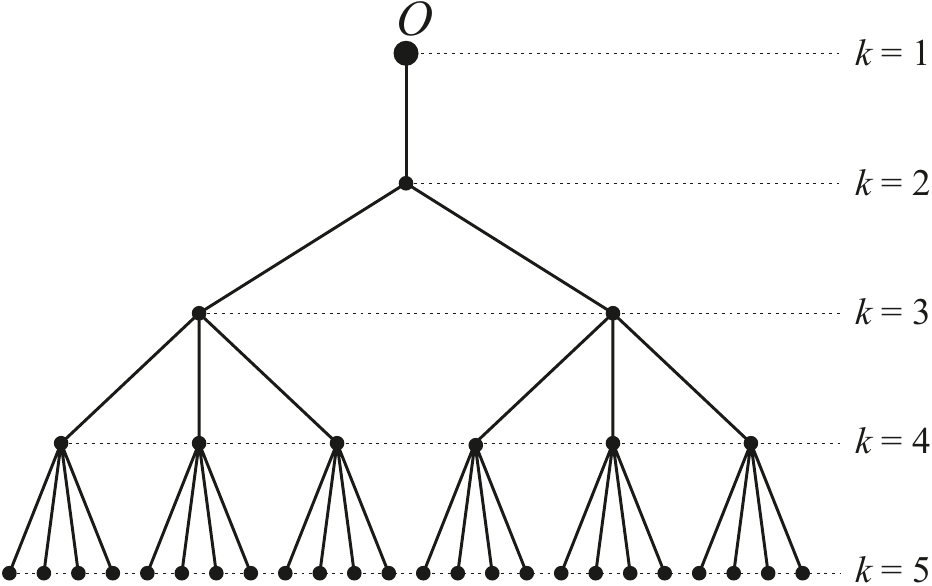}
\caption{A growing tree ${\cal T}$ which has Hermite polynomials as characteristic polynomials.}
\label{fig:03}
\end{figure}

It is eligible to ask the following question: which topology of the target space produces the Hermite polynomials as eigenmodes? This inquiry has been tackled in the works \cite{nech-val-gor1, nech-val-gor2}, wherein it was demonstrated that the Hermite polynomials, $H_k(x)$ serve as the eigenfunctions of a nonuniform growing tree ${\cal T}$. In this tree, the vertex degree is increasing linearly with the distance from the root point as it is shown in \fig{fig:03} where $k$ is the distance from the origin of ${\cal T}$. 

Consider the discrete version of the spectral problem on the graph ${\cal T}$. The matrix $T$ representing ${\cal T}$ is as follows:
\be
T= \left(\begin{array}{cccccc}
0 & 1 & 0 & 0 &   \ldots & 0 \smallskip \\ 1 & 0 & 1 & 0 &  & \smallskip \\
0 & 2 & 0 & 1 &  &  \smallskip \\ 0 & 0 & 3 & 0 &  &  \smallskip \\
\vdots &  &  &  & \ddots &  \smallskip \\ 0 &  &  & & K-1 & 0  \end{array}\right)
\label{eq:20}
\ee
Proceeding in a standard way, we can easily diagonalize the matrix $T$. Introducing the identity matrix $I$, we derive the recursion for characteristic polynomials, $P_k(y)=\det(T-y\, I)$, of the $K\times K$ matrix $T$:
\be
\begin{cases}
P_k(y)=-y P_{k-1}(y)-(k-1)P_{k-2}(y) & \mbox{for $3\le k \le K$} \medskip \\ P_1(y)=y \medskip \\ P_2(y)=y^2-1
\end{cases}
\label{eq:21}
\ee
One can straightforwardly see that $P_k(y)$ coincide with the recursion for the monic Hermite polynomials, ${\cal H}_k(y)$ defined in \eq{eq:10}. Hence, the eigenvalues of the matrix $T$ of size $K\times K$ (see \eq{eq:20}) are the roots of the monic Hermite polynomial, ${\cal H}_k(y)$. We have mentioned already (see \eq{eq:13}) that the asymptotic behavior of monic Hermite polynomials, ${\cal H}_k(\lambda)$ involve the Airy function and at $K\gg 1$ (where $K$ is the maximal size of the matrix ${\cal T}$) the maximal eigenvalue, $y_{max}$, of the transfer matrix \eq{eq:20} has the following asymptotic behavior
\be
y_{max}=2\sqrt{K}+a_1\, K^{-1/6}
\label{eq:22}
\ee
where $a_1\approx -2.3381$ is the first zero of Airy function. Comparing \eq{eq:22} and \eq{eq:14}, derived for stretched orientational diffusion, we can infer that the non-uniformly growing tree ${\cal T}$ serves as a ``spectral target space'' for the orientational diffusion on the surface of the infinite-dimensional hypersphere $S_{\infty}$. Meanwhile, the degree $k$ of the Hermite polynomial $H_k(y)$ represents a ``geodesic distance'' from the root of the tree ${\cal T}$.

The nonuniform tree ${\cal T}$ emerged here as the target space for a discrete Markov process describing orientational diffusion on the surface of a high-dimensional hypersphere. Meanwhile, it was shown in 
\cite{nech-val-gor3} that the tree ${\cal T}$ is essentially a special (and rather atypical) basis for a (growing) Krylov chain \cite{caputa}. The random walk on a growing Krylov chain implies that the particle located at the site $j$ of the Krylov chain, has the weight $\sqrt{j+1}$ for a jump forward and a weight $\sqrt{j}$ for a jump backward -- see \cite{nech-val-gor3} for detail. The Krylov chain, in turn, is tightly linked with the tridiagonalization of Gaussian random matrix ensembles \cite{edelman, caputa}. Thus, one may speculate about a connection between diffusion on the surface of a $D \gg 1$-dimensional hypersphere and the spectral theory of random matrix ensembles.

\bibliography{main-new.bib}

\end{document}